\documentclass{ecai} 
\usepackage{iftex}
\ifpdf
  \usepackage{underscore}         % Only needed if you use pdflatex.
  \usepackage[T1]{fontenc}        % Recommended with pdflatex
\else
  \usepackage{breakurl}           % Not needed if you use pdflatex only.
\fi
\usepackage{times}
\usepackage{soul}
\usepackage{url}
\usepackage[utf8]{inputenc}
\usepackage[small]{caption}
\usepackage{graphicx}
\usepackage{amsmath,amssymb, amsthm}
\usepackage{amsfonts}
\usepackage{booktabs}
\usepackage{algorithm}
\usepackage{algpseudocode}
\usepackage[switch]{lineno}
\usepackage{balance} % for balancing columns on the final page
\usepackage{mathtools}
\usepackage[inkscapearea=page]{svg}
\usepackage{adjustbox}
\usepackage[disable]{todonotes}
\usepackage{subcaption}
\usepackage{centernot}

\usepackage{latexsym}
\usepackage{enumitem}
\usepackage{color}

\newcommand{\BibTeX}{B\kern-.05em{\sc i\kern-.025em b}\kern-.08em\TeX}

\begin{document}

%%%%%%%%%%%%%%%%%%%%%%%%%%%%%%%%%%%%%%%%%%%%%%%%%%%%%%%%%%%%%%%%%%%%%%%%

\begin{frontmatter}

%%% Use this command to specify your submission number.
%%% In doubleblind mode, it will be printed on the first page.

\paperid{123} 

%%% Use this command to specify the title of your paper.

\title{Neural Proofs\\ for Sound Verification and Control of Complex Systems}

\author[A]{\snm{Alessandro Abate}\thanks{Email: alessandro.abate@cs.ox.ac.uk - ORCID id: 0000-0002-5627-9093}}
%\author[B]{\fnms{Second}~\snm{Author}\orcid{....-....-....-....}\footnotemark}
%\author[B,C]{\fnms{Third}~\snm{Author}\orcid{....-....-....-....}} 

\address[A]{OXCAV - Oxford Control and Verification group \\[1ex] 
    Department of Computer Science \\[1ex] 
    University of Oxford, United Kingdom }
%\address[B]{Short Affiliation of Second Author and Third Author}
%\address[C]{Short Alternate Affiliation of Third Author}

\begin{abstract}
This informal contribution presents an ongoing line of research that is pursuing a new approach to the construction of sound proofs for the formal verification and control of complex stochastic models of dynamical systems, of reactive programs and, more generally, of models of Cyber-Physical Systems. 

Neural proofs are made up of two key components:  
1) proof rules encode requirements entailing the verification of general temporal specifications over the models of interest; 
and 2) certificates that discharge such rules, namely they are constructed from said proof rules with an inductive (that is, cyclic, repetitive) approach; 
this inductive approach involves: 
2a) accessing samples from the model's dynamics and accordingly training neural networks, 
whilst 2b) generalising such networks via SAT-modulo-theory (SMT) queries that leverage the full knowledge of the models. 

In the context of sequential decision making problems over complex stochastic models, 
it is possible to additionally generate provably-correct policies/strategies/controllers, namely state-feedback functions that, 
in conjunction with neural certificates, formally attain the given specifications for the models of interest.  
\end{abstract}

\end{frontmatter}

%%%%%%%%%%%%%%%%%%%%%%%%%%%%%%%%%%%%%%%%%%%%%%%%%%%%%%%%%%%%%%%%%%%%%%%%

%%%  MAIN  %%%%%%%%%%%%%%%%%%%%%%%%%%%%%%%%%%%%%%%%%%%%%%%%%%%%%%%%%%%%%%%%%%%%%%%%%%%%

\section{The Complexity of Cyber-Physical Systems} 

The broad area of investigation at OXCAV is that of Cyber-Physical Systems - CPS are complex engineering systems with analogue/physical components, 
the dynamics of which are affected by digital/cyber modules (such as embedded components) \cite{cA17,alurbook,LSAM22}. 
By dynamics we mean the temporal evolution of the memory-full variables (or states) describing the systems of interest over their configuration space (their state space) within an environment.  
Embedded components can additionally comprise data-driven elements, such as modern AI modules that are built from, updated through, or adapted to sensor data. 
CPS thus cover a broad arena of complex systems of interest, spanning across the life and physical sciences, as well as core to many engineering fields.  

\section{Models of Cyber-Physical Systems} 

It is of interest to build dynamical models of CPS that quantitatively/mathematically describe their dynamics across time. 
Such models are classically said to be \textit{hybrid,} encompassing both analogue/continuous and digital/discrete components \cite{alurbook,LSAM22}. 
In view of the complexity of modern CPS, such models can naturally display levels of modularity and hierarchy. 
Ideally, the dynamical models of CPS would be fully mechanistic, namely constructed from first principles and encompassing white-box components, that is with fully-known parts (e.g., parameters).  
More generally, models might be partly known (namely, comprising hidden variables/states that are not fully accessible) and grey-box, whereby parameters are imprecisely known.   
Such uncertainty on model's dynamics and parameters is a staple of CPS models, and can be thought of as a form of epistemic uncertainty - lack of knowledge that, in principle, can be reduced by means of gathering further information about the system, refining (specific parts of) its models or better fitting its parameters and/or estimating its variables/states. 
Additionally and distinctively, uncertainty can manifest itself in aleatoric terms, which usually come as exogenous (external) disturbances that affect the dynamics of the models and that cannot be reduced by means of further observations, or as endogenous (internal) realisations in time of parameters/parts of the model.  
Often (but not strictly necessarily) such uncertainty is modelled by means of endogenous random variables or exogenous stochastic processes, which affect the dynamics of the models of interest.  
Exogenous non-determinism instead can encompass the presence of (non-stochastic) signals originating from the environment, such as disturbances, or alternatively they can be agential. 
Such signals can be selected either adversarially (demonically), or angelically as controllers/strategies/policies. Non-agential non-determinism might be instead resolved under fariness constraints. 
Finally, in the case of the presence of AI modules, such parts of these models are fully black-box, namely known only up to their input-output relations - in practice, this means we can understand them (and possibly further model them) by solely gathering time-series of input and output signals (hence, not leveraging any structural, mechanistic, or a-priori knowledge of their dynamics). These components can be also cast within the broad context of models with partial observations, such as hidden or partially-observable Markov models. 
It is of utmost importance for modern CPS applications to blend in such black-box models within the overall, grey- or white-box models of CPS of interest, as well as to integrate them within the overall analysis of the dynamical properties of said CPS. 
In general, complex CPS often comprise coupled networks of the diverse components we have described (hence the mentioned issues of modularity and hierarchy), 
where the presence of uncertainty (which is lack of determinism that can arise under different shapes and forms, as discussed) needs to be embraced and plays a core role in model development and analysis. 

\section{Model Analysis: Verification and Control Synthesis}  

The considerations above on models naturally lead to discussing issues of model analysis, meaning to mathematically understand a given CPS model across its time evolution.  
More precisely, 
we aim to predict its future evolution (all possible trajectories that can be generated by the model), 
or (whenever possible, and in particular in the presence of agential non-determinism) even to control such future behaviour: 
as John von Neumann is quoted stating, \textit{``all stable systems we shall predict, all unstable we shall control''}.   

By and large, analysis of a model entails studying it under the lens of a specific behavioural property of interest across time - this represents the target of our analysis.   
Basic behavioural properties encompass stability (namely, convergence towards a specific model configuration or a desired trajectory in time), 
transient and steady-state/invariance analysis, 
and robustness against disturbances or uncertainty (whether exogenous or endogenous, whether epistemic or aleatoric). 

In the area of Formal Verification, it is customary - and indeed extremely intuitive, explainable, and technically precise - to express the objective of analysis as a quantitative specification. 
Specifications are intended as formal requirements across time on the dynamical behaviour (the set of all the traces) of the model under study. 
This has been proposed in the context of reactive software systems by \citet{Pnueli77}, who furthered the classical theory of modal logic to be relevant for and applicable to models of software programs.  
This seminal work has spurred the consequent growth of the area of Formal Methods, particularly with research on Model Checking \cite{DBLP:books/daglib/baierkatoen,HandbookMC}, which by now consists of a broad suite of formal approaches and practical software tools for the exhaustive verification of (models of) programs. Whilst we shall not overview this core area in this short contribution, we refer the reader to \citet{DBLP:books/daglib/baierkatoen,DBLP:books/daglib/0007403-2} for a modern presentation of this subject. 
By an large, model checking techniques deal with (models of) systems with finite or countable state spaces (this has notable exceptions, which however encompass very specific and tailored systems, such as timed automata \cite{alurbook,HandbookMC}). 
Instead, we next focus on verification of models of CPS, which (amongst other features) deal with uncountably-infinite (namely dense, if not hybrid) state spaces, stochasticity, and non-linearities -- the presence of these features makes the plug-and-play application of most model checking tools unfeasible in general.  
Extending classical formal verification techniques in these directions is thus relevant and dovetails on the established trend of adoption and extension of notions and approaches from Formal Methods to the area of Control Theory \cite{belta2017formal,tabuada09}, which deals with model that are pertinent to CPSs.    

The behavioural properties of complex dynamical and control models can be investigated from different perspectives and with diverse approaches: 
either analytical (e.g., via local linearisation, evaluation of eigenvalues), or computational ones (e.g., via reach-set computation, by solving optimisation problems, or via formal abstractions as discussed shortly). 
As a general challenge, the non-linearity in the dynamics, the presence of hybrid and of stochastic features in CPS models, 
are difficult to deal with, both analytically and computationally, 
whereas computational approaches (e.g., abstraction-based techniques) are known to be prone to state-space explosion \cite{abate2025archcomp25categoryreportstochastic,LSAM22,SA13}. 

\subsection{Lateral discussion - Formal Abstractions for Verification and Control Synthesis} 

Whilst not the main topic of this contribution, 
we should spend a few lines to describe a relevant, alternative technique for verification and control, 
which is broadly known under the header of  \textit{formal abstractions}.  

The vision underpinning formal abstractions is to quantitatively approximate a given model $\mathfrak S$ with an abstract, computable model $\mathfrak A$. 
Ideally, such abstract model is identical - up to the property under study - to the concrete one; or at least it is in a certain relation to the former \cite{pp09} - in other words, an abstract model isn't just a mere approximation of a given concrete one. 
In practice, this approximation can be made formal and is parameterised by a factor $\epsilon$, 
in that it allows for the definition of a metric $d$ between the models, $d(\mathfrak A, \mathfrak S, \epsilon)$: 
such a metric quantifies the difference between their probability distributions \cite{HSA17},  
or the distance between their solution processes \cite{DBLP:journals/tac/ZamaniEMAL14}.  
We can then adapt the obtained abstraction metric $d$ around the specification of interest $\psi$, 
resulting in an abstract requirement $\psi_{d(\mathfrak A, \mathfrak S, \epsilon)}$. 
The original objective $\mathfrak S \models \psi$ is then sufficiently translated into $\mathfrak A \models \psi_{d(\mathfrak A, \mathfrak S, \epsilon)}$, 
which can now be computed - soundly and automatically - via a model checker from the Formal Verification literature, 
such as PRISM or STORM for probabilistic models, or the mature nuXMV tool for other classes of software programs \cite{HandbookMC}.    
If the outcome is not satisfactory, the procedure ought to allow for the refinement of the abstraction 
into a smaller $\epsilon'$, 
such that $d(\mathfrak A, \mathfrak S, \epsilon') < d(\mathfrak A, \mathfrak S, \epsilon)$. 
This verification approach via formal abstraction can likewise be applied to control synthesis objectives.   

From its seminal developments in \citet{APLS08, AKLP10}, 
the approach has been extended to be applicable to various complex CPS models $\mathfrak S$ \cite{r10}, 
and to rich and diverse properties $\psi$ of interest \cite{LSAM22} - applications cover various domains, such as Autonomy and Smart Energy systems. 
An early prototype software \cite{FAUST15} has been later extended to a plethora of alternative computational tool performing abstraction-based verification and control of stochastic models \cite{abate2025archcomp25categoryreportstochastic}. 
More recently, 
work has placed a particular emphasis has been placed in the quantification and handling of (various forms of) uncertainty: 
for instance, abstraction errors could be embedded within abstract models $\mathfrak A_{d(\mathfrak A, \mathfrak S, \epsilon)}$ as a form of epistemic non-determinism, 
rather than (cf. above) being employed to amend the specification of interest $\psi$ as $\psi_{d(\mathfrak A, \mathfrak S, \epsilon)}$:  
a useful modelling class is that of interval MDPs (iMDPs), as leveraged by the software tool in \citet{StocHy19}.  
This is a very prolific and promising line of research: recently, work has looked at what certificates can be assigned to data-driven abstractions \cite{thom}, 
which are applicable either in the absence of full models, or when it can be solely sampled or simulated.  

As we anticipated above and should be clear by now, one onerous crux of formal abstractions is the scalability to large-scale models. 
Abstraction techniques - whether error- or property-driven, whether lazy or sequential - tend to be costly, either time- or memory-wise: 
state-of-the-art algorithms either slowly generate succinct (smaller) abstractions, 
or more quickly distill expensive abstractions with large state spaces. 
A radically alternative approach to abstraction, which attempts to mitigate these intrinsic shortcomings, is to \textit{learn} such abstract models $\mathfrak A$:  
as spearheaded in \citet{DBLP:conf/nips/AbateEG22}, this sample-based learning can be formally generalised by leveraging (once again) an inductive architecture. 

In conclusion, formal abstraction techniques provide sound and automated results, 
but they should be leveraged only when appropriate. 
On account of their possible limitations, 
alternative approaches can mitigate the stated shortcomings: 
these different approaches provide a (sufficient) proof that the concrete model $\mathfrak S$ fulfils a given requirement $\psi$ in the form of a  \emph{certificate}.  
More precisely, proof certificates offer an approach to verification that avoids the construction of finite abstractions $\mathfrak A$, 
and consequently has the potential to mitigate the state space explosion that is intrinsically associated to them. 
While abstractions construct finite graphs $\mathfrak A$ that are in formal correspondence (as simulations and bisimulations \cite{DBLP:books/daglib/baierkatoen,pp09}) with the original model $\mathfrak S$, and that accordingly enable their analysis via graph algorithms, 
approaches leveraging proof certificates instead seek to construct functions that constitute sufficient proofs for the concrete model $\mathfrak S$ to satisfy a property of interest $\psi$. Such functions, usually mapping the configuration space of the models to an ordered set (e.g., the reals), are what we call \textit{certificates}, and \textit{proofs} (or proof rules) are the requirements that such functions should abide by - requirements that depend on the model's dynamics and on the specification under study. 
The onus of the proof thus shifts to finding, or synthesising, such certificates. 
We shall next be describing how to synthesise \textit{neural certificates} and thus provide what we generally refer to as \textit{neural proofs}.

\section{Neural Proofs}  

Neural proofs are made up of two components:  
1) \textit{proof rules} encode requirements for the verification of general temporal specifications $\psi$ over the models $\mathfrak S$ of interest; 
and 2) corresponding proof \textit{certificates}, which are functions that are constructed from said proof rules, and effectively discharge such proofs. 
The successful synthesis of a certificate soundly asserts that the specification $\psi$ holds on the model $\mathfrak S$. 
This construction can be attained by means of deductive methods, for instance with results in computational algebraic geometry (Farkas' Lemma or Positivstellens\"atze) or via SAT-modulo theory engines, 
or alternatively by inductive (cyclic) approaches, which are the focus of this work.   

In Control Theory, a celebrated instance of such indirect methods is the synthesis of Lyapunov functions \cite{AAGP21,bcAPA20} to prove the (asymptotic) stability of the model dynamics around an equilibrium point: 
these certificates are usually hand-crafted as bespoke energy functions, 
their structure oftentimes built from physics-based consideration of properties and features of the underlying dynamical model $\mathfrak S$. 

In the field of Computer Science (for instance, in the areas of programming languages and theoretical CS) instead, classical certificates are ranking functions or progress measures to study program (or algorithm) termination, 
or invariances (invariant functions) to analyse properties related to program loops. 

We shall explore the state of the art in characterisation and computation of neural proofs for complex CPS models next, 
starting first from the two basic components. 

\subsection{Proof Rules} 

Proof rules formally encode given requirements $\psi$ on the dynamics of the model $\mathfrak S$ under study. 
They allow for the following general formal statement: if a function (the certificate) verifying some properties exists, then the requirements hold in place, and the specification $\psi$ is valid over the model under study $\mathfrak S$, namely we can conclude that $\mathfrak S \models \psi$. 
As such, unsurprisingly the structure of proof rules critically depend on a given model $\mathfrak S$ and specification $\psi$. 

Let us go back to the celebrated examples above. 
For \textit{stability analysis}, a Lyapunov function - which we can intuit as a sort of energy storage - depends on an equilibrium point (around which we study stability) and on a few reasonable conditions (rules): 
it ought to be minimised only at the equilibrium, and it ought to decrease along trajectories around such equilibrium point (namely, if we were to start any execution of the model in a neighbourhood of the equilibrium, and were to lift, or to follow, such trajectory over the surface of the function, then we would see it decreasing along such surface. Clearly, we intuit that such execution would converge to the minimum, namely the equilibrium point itself - hence, we would claim the dynamics are stable around the equilibrium.   

Similarly for for algorithm/program \textit{termination}, a ranking function would be a positive map that admits a finite minimum attained at the states where the program terminates;  
additionally, it would raise a requirement that, 
when evaluated anywhere, 
the function at the next step would always decrease. 
Hence, when existing such a function, 
properly initialised in specific program's configurations, 
would reach its minimum when evaluated at each step of an execution, 
thus proving the termination of the program.  

The reader is now invited to come up with a similar rule defining a function for proving an \textit{invariance} property, along the following lines: 
such a function should uphold a property (e.g., be positive) on a set of states (a putative invariant set), 
and this property should again still hold on all states that are obtained from the transition map applied anywhere within the putative invariant set. 

By now the \textit{leitmotif} should be clear and can be stated generally and in simple terms: 
proof rules are specific algebraic requirements that characterise a certificate, 
which is a function acting on the model's states and upholding certain properties, which depend on a given specification $\psi$, across the dynamics of the model $\mathfrak S$.   
notice that, for a particular property and model's dynamics, 
there may be multiple proof rules - and corresponding certificates - the validity of which sufficiently assert that the property holds on the model of interest. 
It is relevant to come up with non-conservative certificates and, conversely, to show that the existence of a specific certificate is necessary for a property to hold. 

Drawing from the area of Control Theory, 
much work in the literature of certificate synthesis refers to, e.g., \emph{control} Lyapunov functions and \emph{control} barrier certificates \cite{ames2019ControlBarrierFunctions}, 
which have been employed in robotics and control engineering applications. 
It is thus interesting to develop a unified algorithm to concurrently synthesise \emph{both} controllers and certificates in parallel for dynamical models, to prove they satisfy these properties. 
Existing approaches often specify a control set, over which the specifications quantify existentially. In other words, they seek certificates for which there always exists a valid control action that allows for the property to hold. 
A control Lyapunov function therefore proves that there always exists a suitable control input such that the Lyapunov conditions hold, and hence the system is (asymptotically) stabilisable.  
This approach entails that, after a valid control-certificate is synthesised, control actions can be determined,  
e.g. by solving an optimisation program over the input space and the found certificate. 
Instead, \citet{AEP23,cEPA24} synthesise a feedback control law for the model, and apply this state feedback to obtain a \textit{closed-loop} model, for which a certificate is accordingly synthesised. 
These works synthesise the control law \emph{concurrently} with the certificate, hence they verify properties for control models using both a controller and a certificate, and in particular do not delegate the controller synthesis a-posteriori.  
Generalising beyond these standard properties of interest in Control Theory, 
the contribution in \citet{AEP23} summarises a number of properties (or requirements) for dynamical models defined over their trajectories, alongside definitions of corresponding certificates, whose existence serve as \emph{sufficient conditions} for the satisfaction of the desired properties. 
As additionally emphasised in this work, synthesis can in practice and generality be pursued via a data-driven approach, grounded on neural network training (cf. next section). 
The broad class of properties in \citet{AEP23} comprise (finite) compositions of reachability, safety, and invariance requirements, also allowing (as just discussed) for the concurrent synthesis of control policies. 

Conversely, for examples from the area of Computer Science, 
a wealth of research has been thrusted to distill proof rules for the termination of software programs, 
or to prove algorithmic correctness through invariant analysis.  
In particular more recently, for \textit{probabilistic} programs it is of interested to study \textit{qualitative (almost sure) termination}, namely termination with probability one,  
or alternatively \textit{quantitative termination}, namely finding (bounds on) the probability of a program halting \cite{AEGPR23,cav25quantitative,henzinger2025supermartingalecertificatesquantitativeomegaregular}. 
These studies leverage a diverse set of proof certificates that take the form of supermartingales \cite{kallenberg1997foundations}, 
such as ranking \cite{roycav21,TerminationPositivstellensatz}, lexico-graphic \cite{DBLP:journals/corr/abs-1709-04037,DBLP:conf/cav/TakisakaZWL24}, and progress supermartingales \cite{AEGPR23}.  
A number of different, sufficient rules have been found, 
whereas the study of necessary rules (for completeness results) represents a newer and less explored area (cf. Conclusions).   
Next, we examine a few representative proof certificates and related theoretical results from supermartingale theory, 
as classified according to the Manna-Pnueli hierarchy of temporal properties (a more exhaustive overview is given in \citet{DBLP:conf/cav/AbateGR24,cav25quantitative}: 
\begin{itemize}
    \item application of supermartingale convergence to termination/reachability (liveness) specifications \cite[Theorem 4]{DBLP:conf/cav/ChakarovS13}, leveraging \citet[Corollary 4.4.8]{moulinesMarkovChains};  
    
    \item application of supermartingale convergence to recurrence and persistence (liveness) specifications \cite{DBLP:conf/tacas/ChakarovVS16};   
    
    \item application of maximal inequalities to probabilistic safety specifications  
    \cite[Proposition 4.2]{DBLP:journals/toplas/TakisakaOUH21}, leveraging \citet[Corollary 4.4.7]{moulinesMarkovChains};   
    
    \item application of supermartingale convergence and maximal inequalities to reach-avoid specifications \cite[Theorem 1]{DBLP:conf/aaai/ZikelicLHC23} (note that reach-avoidance is a specification that has both a safety component \textit{and} a liveness component);  
    
    \item application of Robbins-Siegmund convergence theorem to $\omega$-regular specifications (including reactivity properties) expressed by deterministic Streett automata: \cite{DBLP:conf/cav/AbateGR24}, \cite[Theorem 10]{cav25quantitative}, and \cite[Figure 5]{DBLP:conf/tacas/DimitrovaFHM16};   
    
    \item application of L{\'e}vy's 0-1 Law to prove completeness (i.e.\ necessary existence) of supermartingale certificates and their use for quantitative probability bounds \cite[Theorem 7, Theorem 8]{cav25quantitative}.
    
\end{itemize}

In summary, as it should be appreciated from the discussion above, 
proof rules entail the existence of a certificate (a function) -  
actually finding or constructing such certificate becomes the practical crux of this general paradigm, which is further elaborated next. 

\subsection{Construction of Proof Certificates} 

We have discussed that, for a broad range of models and properties, different proof rules might work (cf. issue of sufficiency), 
entailing the existence of certificates with specific conditions;  
correspondingly, different algorithms are required to practically compute such specific functions. 
Such algorithms include established and well studied methods based on computational algebraic geometry (including Farkas Lemma, sum-of-squares \cite{parriloThesis} and Positivstellens{\"a}tze \cite{TerminationPositivstellensatz,DBLP:journals/corr/abs-1910-12634}),  
as well as more recent inductive methods that leverage sample-based techniques \cite{AAGP21,roycav21,AEGPR23,DBLP:conf/cav/AbateGR24,bcAPA20,DBLP:conf/atva/AnsaripourCHLZ23,DBLP:conf/tacas/ChatterjeeHLZ23,DBLP:conf/aaai/LechnerZCH22,DBLP:conf/cav/ZhiWLOZ24,DBLP:conf/aaai/ZikelicLHC23,DBLP:conf/nips/ZikelicLVCH23} -   
in this contribution, we shall focus on the latter approach, which is introduced next.   

Inductive synthesis is a general approach that is underpinned by the following philosophy: 
rather than formally and deductively solving a problem, 
one can alternatively guess a solution and then check its validity; 
whenever the check fails, one can additionally garner insight, providing feedback for further guesses whenever the check invalidates the solution. 

\begin{figure}%[t!]
	\centering{
\includegraphics[width=0.3\textwidth]{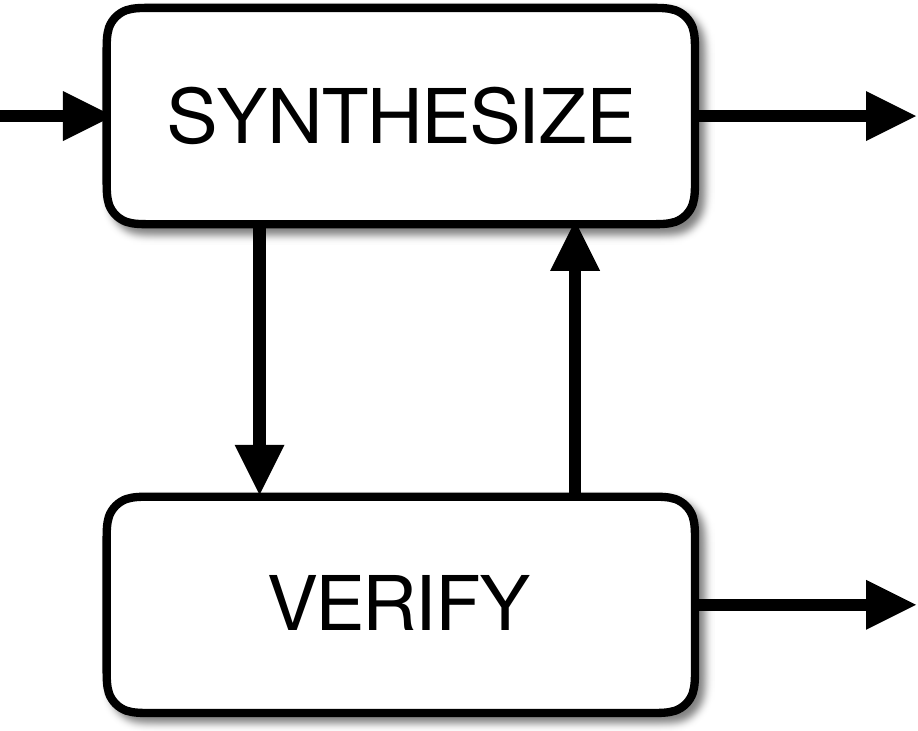}
}
\caption{An inductive synthesis architecture, including a Learner (top block) and a Verifier (bottom), which in particular can be implemented as the known \textit{counterexample-guided inductive synthesis} (CEGIS).}\label{fig:CEGIS} 
\end{figure}

Inductive synthesis has been leveraged across diverse areas and applications, 
and it can be framed in different ways:  
here we focus on a framework known as \textit{counterexample-guided inductive synthesis} (CEGIS), introduced in \citet{solar-lezama2006CombinatorialSketchingFinite}.  
In general, an inductive synthesis framework comprises the interaction of two components (cf. Fig. \ref{fig:CEGIS}): 
a Learner (synthesiser) and a Verifier, namely a module performing an educated guess (synthesis, e.g. via learning), and another one formally checking such guess/hypothesis. 
In principle, the guess can be attempted in numerous-but-incomplete ways, 
for instance, finding parameters of a function (the certificate) over a discrete a finite set of points within the domain of interest. 
Conversely, the Verifier attempts to generalise the hypothesis generated by the Learner, over the entire domain of interest: 
again, this extrapolation (or check) can be attempted in numerous manners, 
and in CEGIS the Verifier soundly validates the candidates obtained from the Learner, by means of calls to a SMT solver that attempts to invalidate the hypothesis.  
SAT-modulo-theory (SMT) extends satisfiability (SAT) solving to richer theories, enabling, for example, finding feasible assignments of real numbered variables over nonlinear formulae \cite{SMT}. 
As such, whenever the candidate is not valid, SMT-generated counter-examples are passed to the Learner for further guessing on an extended set of finite points.  

\subsection{Neural Proofs and Certificates for Models of CPS} 

This inductive synthesis framework is known as CEGIS, 
and despite the admittedly unintuitive interaction between an imprecise Learner and a formal Verifier, 
it surprisingly works on numerous and diverse problems: 
synthesis of Lyapunov functions, 
of barrier certificates, 
and more complex proof certificates. 
Whilst CEGIS has been natively devised for the synthesis of software programs via templates and sketches \cite{DBLP:conf/cav/AbateDKKP18,solar-lezama2006CombinatorialSketchingFinite}, 
more recently it has been employed on CPS studies, 
for the synthesis of digital controllers for continuous plants \cite{ABCCDKKP20}; 
for the hybridisation of nonlinear dynamics for safety verification \cite{DBLP:conf/nips/AbateEG22}; 
and on general applications in real-time autonomy \cite{ADHS24}. 

In neural certificates, 
a central role is placed on te use of neural networks within the CEGIS loop: 
we access samples from the model's dynamics and accordingly training neural networks, 
whilst generalising such networks via SAT-modulo-theory queries that leverage the full knowledge of the models. 
This is motivated by the general approximation capabilities of neural networks, and the recent advances in scalable training of such architectures. 
In particular, the loss function utilised for training is quite intuitive, and simply penalises the deviations (in a certain norm of choice) from the satisfaction of the proof rules above. 
The exchange of information (guesses of putative certificates in one direction, and of counter-example in the opposite one) is important in CEGIS, 
and indeed allows to accommodate additional requirements (progress) or performance goals:  
FOSSIL, a software tool for CEGIS-based synthesis of certificates (and controllers) for general dynamical (and control) models, 
has been released in \citet{cAAEGP21} and extended in \citet{cEPA24}.

\section{Desiderata for Future Work on Neural Proofs}\label{sec:conclusion} 

That of neural proofs is a blossoming research area, 
with contributions that promise to be impactful in diverse control engineering applications, 
for the formal verification of software programs, 
and as a proof engine to prove properties of practical algorithms. 
In particular for CPS applications, there is an acute need for scalable solutions - eg based on SMT \cite{SMT} - that aptly and practically harness the complexity of dense and infinite-state and -action models.  
In order to attain scalability and independence from given models of systems under study, 
\citet{rickard2025datadrivencertificatesynthesis} eschews the use of SMT in the Verifier block of Figure \ref{fig:CEGIS} and leverages instead the very samples employed to train the certificate within the Learner to provide statistical guarantees on the obtained function. Such statistical assertions follow the PAC (probably approximately correct) theory. 

For stochastic models of CPS, further work on quantitative supermartingale certificates are much needed for ever more general specifications, building on the recent contributions in \cite{cav25quantitative,henzinger2025supermartingalecertificatesquantitativeomegaregular} that move beyond the class of finite-horizon (safe and co-safe) properties in \citet{AEP23}. 
Ongoing work on tightening sufficiency results towards completeness \cite{DBLP:conf/popl/FioritiH15,majumdar2023positivealmostsuretermination} is also of theoretical interest. 

Finally, algorithms and well maintained software tools for the practical computation of certificates, moving beyond \cite{cAAEGP21,cEPA24}, 
are surely needed and of potential impact.   

%%%%%%%%%%%%%%%%%%%%%%%%%%%%%%%%%%%%%%%%%%%%%%%%%%%%%%%%%%%%%%%%%%%%%%%%

\begin{ack}
I wish to thank long-term collaborators contributing to work on certificates and abstractions, 
well as present and past members of the OXCAV group, 
in particular D. Roy, A. Edwards, L. Rickard, M. Giacobbe, E. Polgreen (certificates), and 
T. Badings, N. Vertovec, V. Debauche, A. Su\~n\'e, M. Nazeri, L. Romao, F. Cosentino, N. Cauchi, I. Tkachev, S. Haesaert, S. Soudjani, M. Zamani (abstractions). 
\end{ack}

%%%%%%%%%%%%%%%%%%%%%%%%%%%%%%%%%%%%%%%%%%%%%%%%%%%%%%%%%%%%%%%%%%%%%%%%

\newpage 

\bibliography{references}

\end{document}